\documentclass[runningheads]{llncs}
\usepackage[T1]{fontenc}
\usepackage{graphicx,verbatim}
\usepackage[T1]{fontenc}
\usepackage{graphicx}
\usepackage{diagbox}
\usepackage{amsfonts}
\usepackage{multirow}
\usepackage{enumitem}
\usepackage{amsmath}

\usepackage{hyperref}
\usepackage[capitalise]{cleveref}
\usepackage{xcolor,soul}
\hypersetup{
    colorlinks=true,
    linkcolor=blue,
    filecolor=cyan,      
    urlcolor=magenta,
}
\usepackage{booktabs}
\usepackage{color}

\usepackage{bbm}
\usepackage{float}
\usepackage{color, colortbl}
\usepackage{amssymb}  
\usepackage{pifont}   
\usepackage{subcaption} 
\usepackage{tikz}  
\begin{document}
\title{LHU-Net: a Lean Hybrid U-Net for Cost-efficient, High-performance Volumetric Segmentation}
\author{Yousef Sadegheih $^\dag$\inst{1} \and
Afshin Bozorgpour $^\dag$\inst{1} \and
Pratibha Kumari\inst{1} \and
Reza Azad\inst{2}\and
Dorit Merhof\inst{1,3}}


%
\authorrunning{Y. Sadegheih et al.}
%
\institute{Faculty of Informatics and Data Science, University of Regensburg, Regensburg, 93053, Germany \and
Faculty of Electrical Engineering and Information Technology, RWTH Aachen University, 52062 Aachen, Germany \and
Fraunhofer Institute for Digital Medicine MEVIS, Bremen 28359, Germany
\email{dorit.merhof@ur.de}\\
$\dag${\small \textit{ Indicates equal contribution}}}
    
\maketitle              
\begin{abstract}
The rise of Transformer architectures has advanced medical image segmentation, leading to hybrid models that combine Convolutional Neural Networks (CNNs) and Transformers. However, these models often suffer from excessive complexity and fail to effectively integrate spatial and channel features, crucial for precise segmentation. To address this, we propose LHU-Net, a Lean Hybrid U-Net for volumetric medical image segmentation. LHU-Net prioritizes spatial feature extraction before refining channel features, optimizing both efficiency and accuracy. Evaluated on four benchmark datasets (Synapse, Left Atrial, BraTS-Decathlon, and Lung-Decathlon), LHU-Net consistently outperforms existing models across diverse modalities (CT/MRI) and output configurations. It achieves state-of-the-art Dice scores while using four times fewer parameters and 20\% fewer FLOPs than competing models, without the need for pre-training, additional data, or model ensembles. With an average of 11 million parameters, LHU-Net sets a new benchmark for computational efficiency and segmentation accuracy. Our implementation is available on \href{https://github.com/xmindflow/LHUNet}{github.com/xmindflow/LHUNet}.


\end{abstract}

\section{Introduction}\label{sec:intro}


        



Medical image acquisition technologies such as MRI, CT, and X-ray enable non-invasive imaging of anatomical structures, making image segmentation essential for diagnosis, intervention planning, and disease assessment. Manual segmentation is time-consuming and prone to inconsistencies, necessitating automated methods. While deep learning approaches, particularly Convolutional Neural Networks (CNNs), have advanced medical segmentation, their performance can be limited by a lack of global context~\cite{azad2024medical,isensee2021nnu}. Vision Transformers (ViTs)~\cite{geirhos2018imagenet}, which use self-attention to capture global context, have addressed this gap but often fail to preserve fine-grained local details essential for accurate segmentation~\cite{azad2024advances}. 
\begin{figure*}[!t]
\resizebox{\textwidth}{!}{
        \begin{tabular}{@{} ccccc @{}}
               \includegraphics[width=0.1\textwidth]{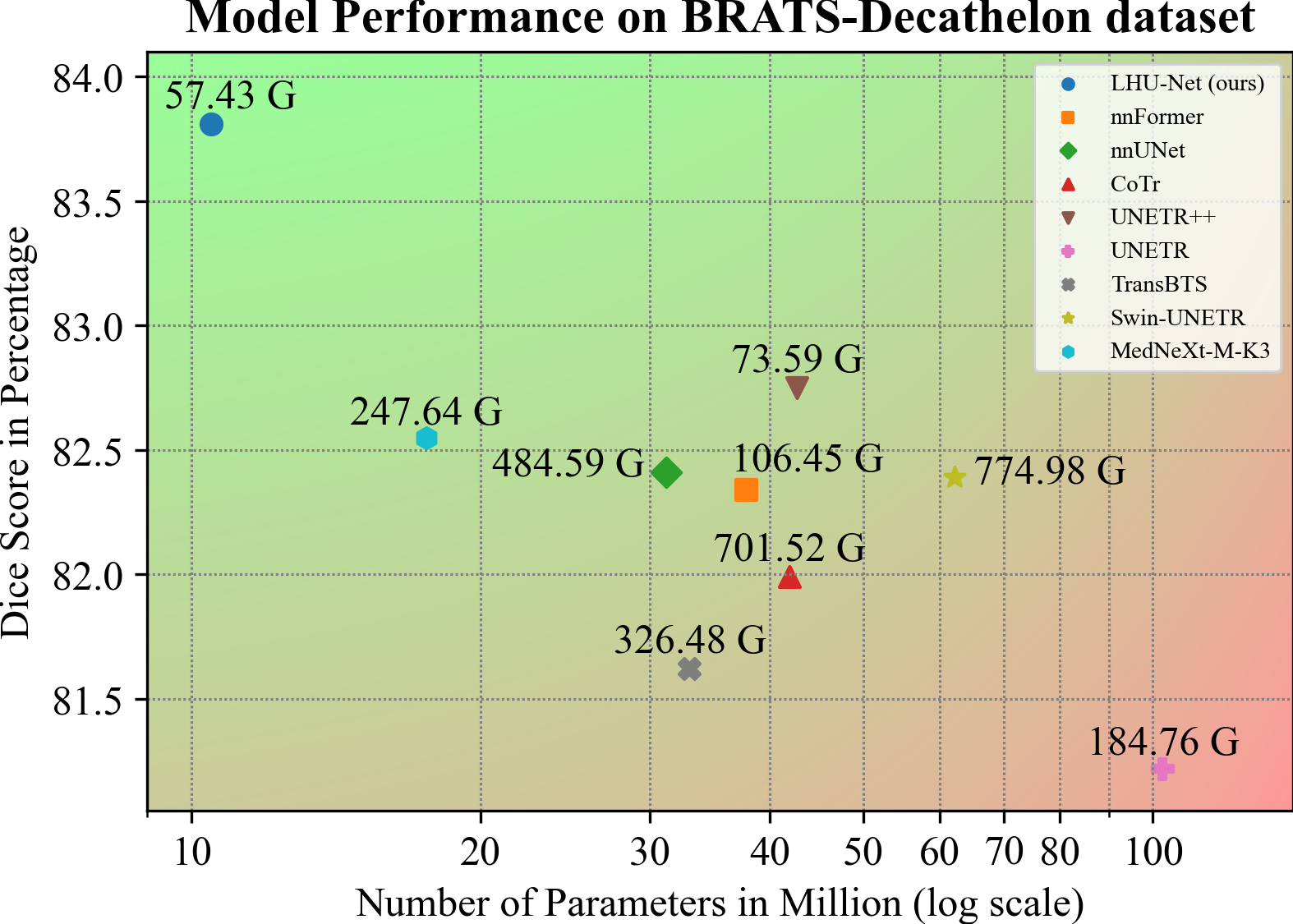}  & & & & 
                \includegraphics[width=0.105\textwidth]{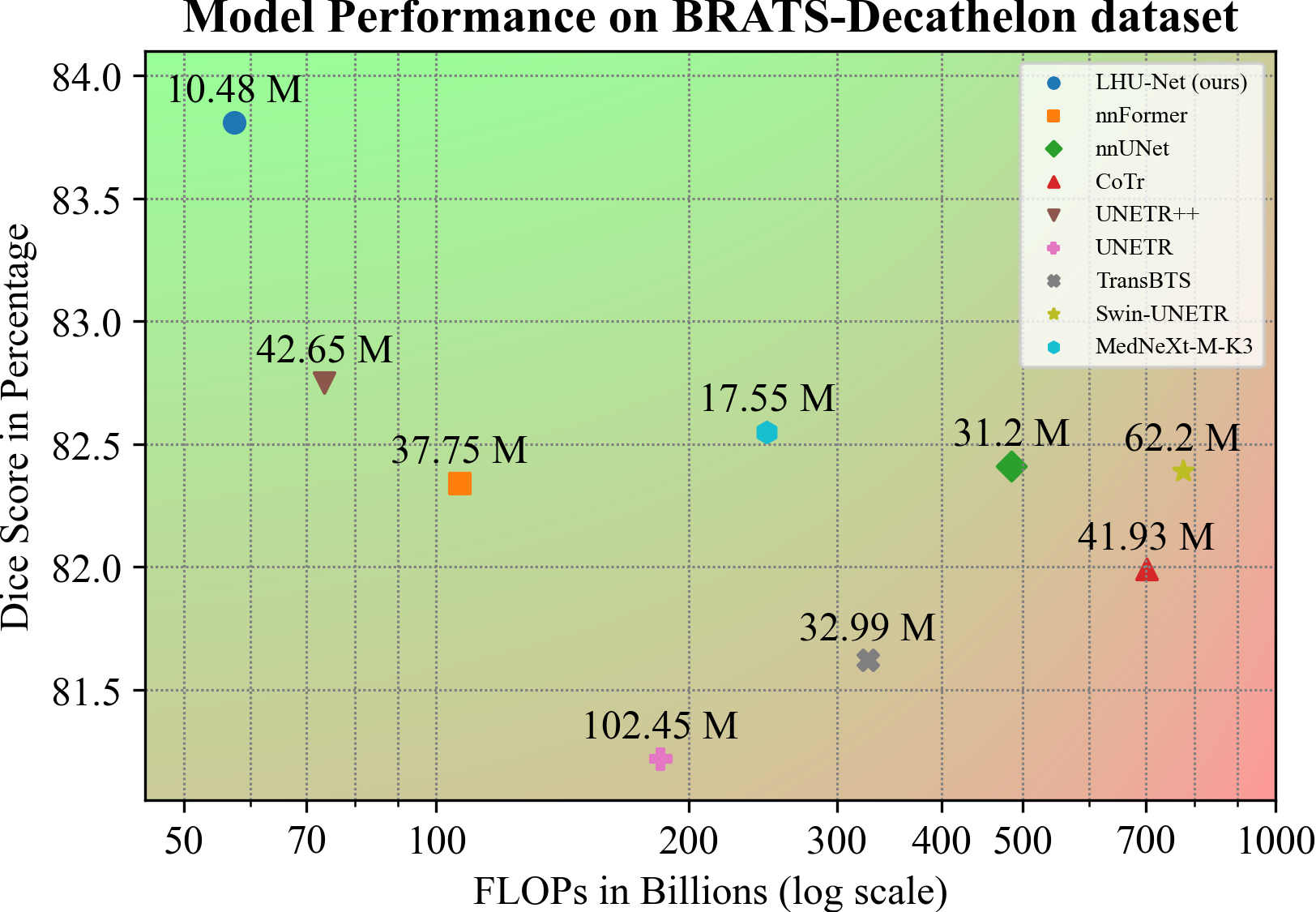}
        \end{tabular}
    }
    \caption{Model performance on the BraTS-Decathlon dataset: (left)  DSC vs. parameter count and (right) DSC vs FLOPs.}
    \label{fig:combined_performance}
 \end{figure*}
Segmentation becomes even more challenging in the 3D domain due to the increased data volume and complexity. However, 3D models have been shown to outperform 2D models by capturing better context and improving segmentation accuracy. Despite this, 3D models typically require higher computational power and parameter counts~\cite{azad2024beyond,hatamizadeh2022unetr}. A common trend is to use the same module across all layers to achieve state-of-the-art performance~\cite{azad2024beyond,yang2024hca,chen2023collaborative,shaker2022unetr++}. However, we argue that using tailored modules for different layers can make the model more efficient, achieving better segmentation with lower computational costs.
Hybrid models, which combine CNNs for local feature extraction with ViTs for global context, have gained popularity in tackling these challenges. While promising, many existing hybrid models increase complexity without proportional improvements in performance, leading to excessive computational costs~\cite{zhou2021nnformer,he2023swinunetr,hatamizadeh2021swin}. To address this, LHU-Net (\textbf{L}ean \textbf{H}ybrid \textbf{U-Net}) optimizes attention mechanisms by using spatial attention in early layers for local feature extraction and channel attention in deeper layers for broader contextual understanding. This approach balances model complexity with efficiency, significantly reducing computational cost while improving 3D medical image segmentation performance. In this paper, we present LHU-Net with the
following contributions: \ding{182} \textbf{Efficient Hybrid Attention selection for Better Contextual Understanding:} LHU-Net utilizes two specialized attention mechanisms within ViTs to capture both local and global contexts effectively. It combines Large Kernel Attention with an extra deformable attention layer to manage long-range dependencies and maintain high-frequency details. In the early layers, spatial attention focuses on local features, while in deeper layers, channel attention captures global feature interactions, ensuring a comprehensive feature extraction process suited for medical image segmentation. \ding{183} \textbf{High Efficiency with Minimal Cost:} On BraTS, it reduces parameters by 75\% and FLOPS by 21\% while maintaining top DSC performance, averaging 11 M parameters across datasets (\Cref{fig:combined_performance}). \ding{184} \textbf{Robust Across Modalities:} Excelling in CT, MR, and multimodal datasets, LHU-Net handles both single- and multi-label segmentation tasks with high versatility.

\begin{figure*}[!ht]
\centering
\includegraphics[scale=0.45]{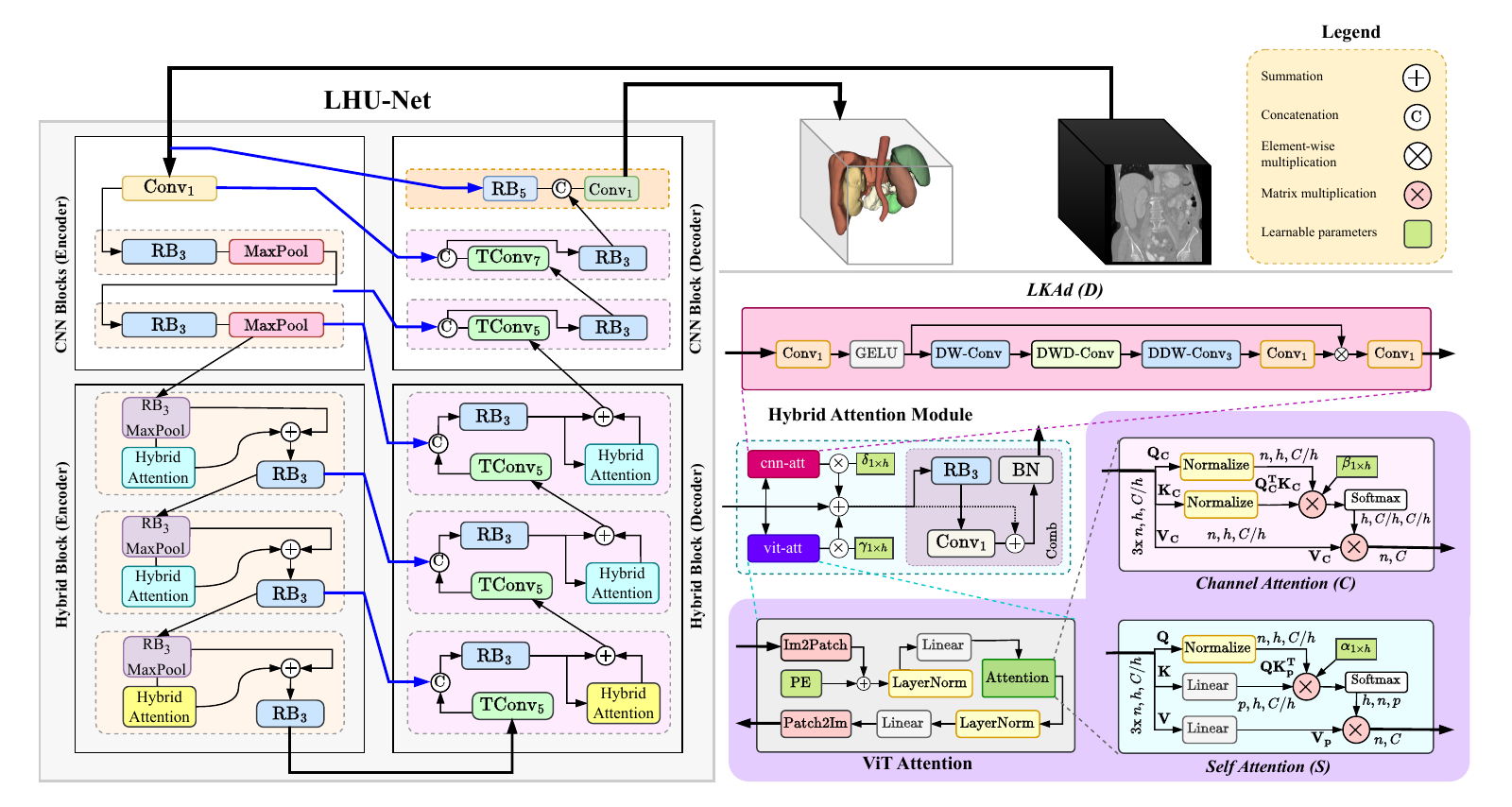}
\caption{Overview of the LHU-Net architecture.}

\label{fig:network}
\end{figure*}

\section{Methodology}\label{sec:method}
LHU-Net extracts features efficiently by combining convolutional blocks with hybrid attention mechanisms to capture both local and global contexts. As shown in~\Cref{fig:network}, its U-Net encoder-decoder processes 3D patches, with the encoder refining multi-scale features and skip connections transferring key details for segmentation reconstruction. Initial convolutional blocks enhance local features and reduce spatial dimensions, while core hybrid blocks integrate large kernel convolutional attention followed by deformable convolution and spatial-channel ViT attention to capture global features and long-range dependencies. Detailed explanations of each stage follow.

\subsection{CNN Blocks}
Our architecture employs CNN blocks to capture local features at high and mid-frequency levels efficiently. Unlike models relying solely on ViTs, we use ResBlocks with MaxPool to downsample spatial dimensions while preserving key details for later hybrid processing. Before the encoder, an initial refinement is applied using a point-wise convolution (PW-Conv) followed by PReLU and batch normalization to maintain spatial resolution and sharpen boundaries. Each encoder layer employs a ResBlock that begins with a depth-wise convolution, batch normalization, and leakyReLU, followed by another depth-wise convolution with batch normalization. A residual connection, processed through PW-Conv and batch normalization, is added before a final leakyReLU activation. MaxPool is then used to downsample the spatial dimensions, which reduces the feature map size and aggregates prominent features, making the subsequent processing more effective. This approach achieves two key objectives. It reduces the computational cost of processing large spatial dimensions with ViT blocks and preserves local features by delaying downsampling until after initial refinement, thereby optimizing parameter usage.


\begin{table}[t!]
\renewcommand{\r}[1]{\textcolor{red}{#1}}
\renewcommand{\b}[1]{\textcolor{blue}{#1}}
\centering
\caption{LHU-Net training configuration along with parameter and FLOPs comparisons wrt SOTA models. \b{Blue} and \r{red} indicate the best and second-best results, respectively.}
\resizebox{\textwidth}{!}{
\begin{tabular}{l||c|c||c|c||c|c||c|c}
\toprule
\textbf{LHU-Net Details}   & \multicolumn{2}{c||}{\textbf{Synapse}}                   & \multicolumn{2}{c||}{\textbf{Lung-Decathelon}}                                  & \multicolumn{2}{c||}{\textbf{BraTS-Decathelon}}           & \multicolumn{2}{c}{\textbf{LA}} \\
\toprule
Patch size         & \multicolumn{2}{c||}{128 x 128 x 64}            & \multicolumn{2}{c||}{192 x 192 x 32}                        & \multicolumn{2}{c||}{128 x 128 x 128} & \multicolumn{2}{c}{96 x 96 x 96}   \\
Base learning rate & \multicolumn{2}{c||}{0.003}                     & \multicolumn{2}{c||}{0.003}                                 & \multicolumn{2}{c||}{0.01}            & \multicolumn{2}{c}{0.01}           \\
Downsample\footnotemark        & \multicolumn{2}{c||}{{[}2, 2, 1{]}, 2, 2, 2, 2} & \multicolumn{2}{c||}{{[}2, 2, 1{]}, {[}2, 2, 1{]}, 2, 2, 2} & \multicolumn{2}{c||}{2, 2, 2, 2, 2}   & \multicolumn{2}{c}{2, 2, 2, 2, 2}  \\
\toprule

\textbf{Methods}   & \textbf{Params.\(\downarrow\)}            & \textbf{FLOPs\(\downarrow\)}           & \textbf{Params.\(\downarrow\)}                   & \textbf{FLOPs\(\downarrow\)}                 & \textbf{Params.\(\downarrow\)}       & \textbf{FLOPs\(\downarrow\)}       & \textbf{Params.\(\downarrow\)}       & \textbf{FLOPs\(\downarrow\)}     \\
\toprule

nnUNet~\cite{isensee2021nnu} & 30.71 M  & 476.58 G &30.79 M&561.82 G & 31.20 M & 484.59 G &30.79 M & 539.34 G\\
\hline

MedNeXt-M-K3~\cite{roy2023mednext} &\r{17.55 M}  & 123.47 G &\r{17.55 M} & 137.48 G & \r{17.55 M} & 247.64 G &\r{17.55 M} & 103.84 G \\

MedNeXt-M-K5~\cite{roy2023mednext} & 18.26 M  & 153.52 G &18.26 M & 171.29 G & 18.26 M  & 307.74 G &18.26 M & 129.20 G \\

CoTr~\cite{xie2021cotr} & 41.87 M  &334.23 G & 41.86 M & 375.11 G & 41.93 M & 701.52 G &41.86 M & 281.33 G \\

nnFormer~\cite{zhou2021nnformer} &150.5 M   &213.4 G & 149.12 M &136.10 G & 37.75 M & 106.45 G &149.22 M & 102.35 G \\

UNETR++~\cite{shaker2022unetr++} &42.96 M   &\r{47.98 G} & 121.18 M & 125.84 G & 42.65 M &\r{73.59 G} &29.54 M &\r{29.74 G} \\

Swin-UNETR~\cite{hatamizadeh2021swin} & 62.83 M  &384.2 G & 62.19 M& 429.95 G & 62.2 M & 774.98 G &62.19 M & 319.38 G \\

TransBTS~\cite{wenxuan2021transbts} & 32.79 M &324.13 G&32.79 M &323.73 G &32.99 M  &326.48 G  &31.58 M & 119.81 G \\

UNETR~\cite{hatamizadeh2022unetr} &92.49 M   &75.76 G & 121.18 M& \r{98.40 G} & 102.45 M & 184.76 G &92.78 M & 73.51 G \\

\midrule
\rowcolor[HTML]{C8FFFD}
\textbf{LHU-Net} &\b{10.47 M}   & \b{37.49 G }&\b{14.68 M}& \b{52.26 G} & \b{10.48 M} &\b{ 57.43 G }&\b{8.53 M} &\b{ 22.20 G} \\
\bottomrule
\end{tabular}
}
\label{tab:experimets-config}
\end{table}
\footnotetext{A single integer indicates uniform downsampling across all axes.}

\subsection{Hybrid Blocks (Hybrid Attention)}
In the intermediate and deep layers, we employ hybrid blocks to enhance segmentation by balancing local detail with global context. By integrating the LKAd module as a CNN attention block, these blocks improve boundary delineation and object identification. Building on robust local features from earlier stages, our hybrid blocks use larger kernel sizes to capture broader spatial representations and deploy deformable attention to focus on relevant receptive fields. Spatial ViT attention is applied in the early part of this stage, while channel ViT attention in later layers aggregates high-level features. This design overcomes the limitations of repeated fixed modules~\cite{shaker2022unetr++,rahman2024mist,liu2024scanext} and avoids the redundancy seen in methods that repeat blocks across levels~\cite{azad2024beyond,wang2023cross,hong2023dual,li2023cpftransformer}. Our level-specific modules ensure that each stage effectively captures its unique features, resulting in improved segmentation performance with fewer parameters and lower computational cost.

\textbf{Self-Adaptive Contextual Fusion Module:} This module is integrated into the top hybrid blocks to enhance global structure capture by fusing spatial attention maps from both the LKAd (D) and self-attention (S) mechanisms. This effectively preserves global context and minimizes information loss in deep encoder layers. Within the module, LKAd uses a deformable grid to adaptively capture local and global features, while S handles long-range dependencies and reduces texture bias to maintain high-frequency details. The final output is computed as:
\begin{equation}
F_S = Comb\left(F + \delta_s\, \mathbf{D}(F) + \gamma_s\, \mathbf{S}(F)\right)
\label{eq:FS}
\end{equation}
where \( F \in \mathbb{R}^{C \times H \times W \times D} \) is the input feature map, \( F_S \) is the output, and \(\delta_s\) and \(\gamma_s\) are learnable weights that balance the contributions of the two attention mechanisms. The $Comb$ function, as shown in \Cref{fig:network}, processes the fused output through a ResBlock and a PW-Conv, followed by a residual connection and batch normalization. This design ensures effective feature fusion and robust contextual representation.

\begin{table}[t!]
\renewcommand{\r}[1]{\textcolor{red}{#1}}
\renewcommand{\b}[1]{\textcolor{blue}{#1}}
\centering
\caption{Comparison of LHU-Net with SOTA methods on the Synapse dataset. \b{Blue} and \r{red} indicate the best and second-best results, respectively. Metrics include DSC for individual organs, average DSC, and HD95. Parameter counts (M) and FLOPs (G) are also reported.}
\resizebox{\textwidth}{!}{
\begin{tabular}{l||cc||cccccccc||cc}
\toprule
\multirow{2}{*}{\bf{Methods}} &   \multirow{2}{*}{\bf{Params}\(\downarrow\)}   & \multirow{2}{*}{\bf{FLOPS}\(\downarrow\)} & \multirow{2}{*}{\bf{Spl}} & \multirow{2}{*}{\bf{RKid}} & \multirow{2}{*}{\bf{LKid}} & \multirow{2}{*}{\bf{Gal}} & \multirow{2}{*}{\bf{Liv}} & \multirow{2}{*}{\bf{Sto}} & \multirow{2}{*}{\bf{Aor}} & \multirow{2}{*}{\bf{Pan}} & \multicolumn{2}{c}{\bf{Average}}\\ 
\cline{12-13} &  & & & & & & & & & & \bf{DSC} \(\uparrow\) & \bf{HD95} \(\downarrow\)\\ 
\midrule
UNETR~\cite{hatamizadeh2022unetr}	                    & 92.49 M           & 75.76 G       & 85.00         & 84.52           & 85.60        & 56.30        & 94.57         & 70.46        & 89.80          & 60.47        & 78.35         & 18.59 \\

Swin-UNETR~\cite{hatamizadeh2021swin}	                & 62.83 M           & 384.2 G           & 95.37         & 86.26           & 86.99       & 66.54       & 95.72         & 77.01        & 91.12         & 68.80         & 83.48         & 10.55 \\

nnFormer~\cite{zhou2021nnformer}	                    & 150.5 M           & 213.4 G           & 90.51         & 86.25           & 86.57       & 70.17       & 96.84     & \b{86.83}    & 92.04         & \b{83.35}    & 86.57         & 10.63 \\

UNETR++~\cite{shaker2022unetr++} 	                    & 42.96 M           & \r{47.98 G}       & 95.77     & 87.18           & 87.54       & 71.25       & 96.42         & \r{86.01}        & \r{92.52}     & 81.10         & 87.22    & \r{7.53} \\


TC-CoNet~\cite{chen2023collaborative}                   & 313.67 M          & 699.58 G          & 91.77         & \r{87.92}       & \b{88.16}       & 62.00       & 96.35         & 79.40         & 92.45     & 72.78        & 83.86         & 9.64 \\ 

D-LKA Net~\cite{azad2024beyond}
& 42.35 M           & 66.96 G           & \r{95.88}         & \b{88.50}   & 87.64  & 72.14  & 96.25   & 85.03   & \b{92.87}   &  81.64  &      \r{87.49}   &  9.57 \\



TransBTS~\cite{wenxuan2021transbts} &32.79 M &324.13 G& 91.65 & 86.99 & 87.46 & 62.52 & 96.42 & 77.39 & 91.71 & 72.12 & 83.28 & 12.34  \\

CoTr~\cite{xie2021cotr} &41.87 M  &334.23 G& 94.93 & 86.80 & 87.67 & 62.90 & 96.37 & 80.46 & 92.43 & 78.84 & 85.05  & 9.04  \\

nnUNet~\cite{isensee2021nnu} &30.71 M  & 476.58 G& 91.16 & 86.21 & 86.92 & 69.77 & 96.49 & 85.92 & 91.78 & \r{83.23} & 86.44  & 10.91  \\

MedNeXt-M-K3~\cite{roy2023mednext} &\r{17.55 M}  & 123.47 G& 90.63 & 86.50 & 87.66 &\r{73.00} & \r{96.92} & 77.89 & 92.25 & 80.81 & 85.71  & 19.10  \\

MedNeXt-M-K5~\cite{roy2023mednext} &18.26 M  & 153.52 G& 91.16 & 87.51 & 87.67 & 71.31 &\b{97.01} & 80.46 & 92.48 & 80.20 & 85.97 & 17.59  \\


\midrule
\rowcolor[HTML]{C8FFFD}
\bf{\textit{LHU-Net}}                                            & \b{10.47 M}       & \b{37.49 G}       & \b{96.02}     & 87.46       & \r{87.75}   & \b{74.30}	      & 96.83     & 85.73        & \r{92.53}         & 82.04        & \b{87.83}     & \b{6.26} \\

\bottomrule
\end{tabular}
}

\label{tab:results-synapse}
\end{table}

\textbf{LKAd (D)}: This module computes attention following the method in~\cite{azad2024beyond}. The input tensor \(x\in\mathbb{R}^{C\times H\times W\times D}\) undergoes these steps:
\begin{equation}
\begin{aligned}
\hat{x} &= \text{GELU}(\text{PW-Conv}(x)),\\
x_{\text{LKAd}} &= \text{PW-Conv}(\text{DDW-Conv}_3(\text{DWD-Conv}\bigl(\text{DW-Conv}(\hat{x}))))\otimes \hat{x},\\
\end{aligned}
\label{eq:dlka}
\end{equation}

First, \(x\) is refined via a pointwise convolution and GELU activation, producing \(\hat{x}\). Next, sequential depth-wise and dilated convolutions extract multi-scale local features, The key step is the deformable depth-wise convolution (\(\text{DDW-Conv}_3\)), which enables adaptive sampling, enhancing fine-detail and long-range dependency capture. Finally, element-wise multiplication with \(\hat{x}\) yields \(x_{\text{LKAd}}\), a rich, contextually enhanced representation of \(x\).

\textbf{Self-Attention (S)}: The self-attention module computes \( S(x) \) from a normalized tensor \( x \) of shape \( n \times C \), where \( n = H \times W \times D \). First, three linear layers generate queries, keys, and values as \( Q = W^Q x,\; K = W^K x,\; V = W^V x \), each reshaped to \( n \times h \times \frac{C}{h} \) for \( h \) heads. Following~\cite{shaker2022unetr++}, two additional projections transform \( K \) and \( V \) into learnable representations:
\(K_p = W^p_K K,\; V_p = W^p_V V,\)
reducing the spatial dimension from \( n \) to \( p \) (with \( p \ll n \)),  which improves efficiency and focuses attention on the most representative features. Next, the normalized \( Q \) is multiplied by \( K_p^T \) (scaled by \(\sqrt{d}\)) and the resulting values are adjusted using the learnable parameter $\alpha$ and passed through a softmax ($\sigma$)to obtain similarity scores, which are then multiplied by \( V_p \) to yield the spatial attention where it is further refined by a linear normalization and a linear layer. 

\textbf{OmniFocus Attention Block:} At the deepest network level, this block processes the richest feature layers by operating residually in the encoder and collectively in the decoder. It enhances feature representation by reducing noise and capturing essential details through integrated convolutional flows. The block leverages ViT channel attention (C) alongside the LKAd module to learn inter-channel relationships and long-range dependencies. Specifically, the block applies the LKAd module (see Equation~\ref{eq:dlka}) and a channel attention module that computes inter-channel interactions using dot-product attention on projections \(Q_C\), \(K_C\), and \(V_C\):
\begin{equation}
\begin{aligned}
x_{C} = V_C \cdot \sigma\left( \frac{Q_C^T K_C}{\sqrt{d}} \right).
\end{aligned}
\label{eq:attention}
\end{equation}
After linear normalization and an additional linear layer, the final output is produced by a $Comb$ function that fuses the LKAd and channel attention outputs with learnable weights (similar to the Self-Adaptive Contextual Fusion Module). This design avoids redundant module repetition and effectively extracts rich contextual information, thereby enhancing overall segmentation performance.

\begin{table*}[!t]
\renewcommand{\r}[1]{\textcolor{red}{#1}}
\renewcommand{\b}[1]{\textcolor{blue}{#1}}
\centering
\caption{Comparison of LHU-Net with SOTA models on BraTS, Lung, and LA datasets. \b{Blue} and \r{red} indicate the best and second-best results, respectively.}
\small
\resizebox{\textwidth}{!}{
\begin{tabular}{l||ccc|cc||cc||cc}
\toprule
\multirow{2}{*}{\textbf{Method}} & \multicolumn{5}{c||}{\textbf{BRATS (MRI)}} & \multicolumn{2}{c||}{\textbf{Lung (CT)}} & \multicolumn{2}{c}{\textbf{LA (MRI)}}  \\
\cmidrule(lr){2-6} \cmidrule(lr){7-8} \cmidrule(lr){9-10} 
& WT & ET & TC & DSC $\uparrow$ & HD95 $\downarrow$ & DSC $\uparrow$ & HD95 $\downarrow$ & DSC $\uparrow$ & HD95 $\downarrow$ \\
\midrule
UNETR~\cite{hatamizadeh2022unetr} & 90.35 & 76.30 & 77.02 & 81.22 & 6.61 & 73.29 &  23.84 & 91.25 & 9.23 \\
TransBTS~\cite{wenxuan2021transbts} & 90.91 & 77.86 & 76.10 & 81.62 & 5.80 & 70.38 &  30.09 & 92.25 & 4.92\\
Swin-UNETR~\cite{hatamizadeh2021swin} & 91.12 & 77.65 & 78.41 & 82.39 & 5.33 & 75.55 & 28.74 & 91.89 & 5.96\\
CoTr~\cite{xie2021cotr} & 91.01 & 77.52 & 77.43 & 81.99 & 5.78 & 75.74 & 27.91 & 92.60 & 4.87  \\
nnUNet~\cite{isensee2021nnu} & 91.21 & 77.96 & 78.05 & 82.41 & 5.58 & 74.31 & 28.52 & \r{92.75} & \r{4.35} \\
nnFormer~\cite{zhou2021nnformer} & 91.23 & 77.84 & 77.91 & 82.34 & 5.18 & 77.95 & 16.25 & 92.02 & 5.08 \\
MedNeXt-M-K3~\cite{roy2023mednext} & \r{91.42} & 78.24 & 77.98 & 82.55 & 5.13 & 80.14 &2.85 & 92.68 & 4.49  \\
MedNeXt-M-K5~\cite{roy2023mednext} & 91.21 & 78.15 & 78.03 & 82.46 & 5.37 & 79.51 &\r{2.84} &92.50 & 4.68\\
UNETR++~\cite{shaker2022unetr++} & 91.27 & \r{78.39} & \r{78.60} & \r{82.75} & \r{5.05} &\r{80.68} &\b{2.79} & 92.55 & 5.08  \\
\midrule
\rowcolor[HTML]{C8FFFD}
\bf{LHU-Net}                                & \b{91.56}         & \b{80.03}     & \b{79.83}      & \b{83.81}     & \b{4.83} & \b{81.27 }& 3.23 & \b{92.91}& \b{3.99}   \\ 
\bottomrule
\end{tabular}
}

\label{tab:results-LA}
\end{table*}

\section{Experimental setup and results}\label{sec:results}
\subsection{Datasets, experimental setup and evaluation metrics}
To assess model effectiveness, we used four datasets: two CT-based and two MRI-based. \textbf{Synapse}~\cite{landman2015miccai} consists of 30 abdominal CT scans, split as in~\cite{azad2024beyond,shaker2022unetr++} (18 for training, 12 for testing). Evaluated on eight organs: aorta, gallbladder, spleen, left/right kidneys, liver, pancreas, and stomach. \textbf{Lung-Decathlon}~\cite{antonelli2022medical,simpson2019large} consists of 63 CT scans for lung cancer segmentation, using an 80:20 training-validation split~\cite{shaker2022unetr++}. \textbf{BraTs-Decathlon}~\cite{antonelli2022medical,menze2014multimodal} which consists of 484 MRI scans (FLAIR, T2w, T1w, T1ce) for segmenting whole tumor (WT), enhancing tumor (ET), and tumor core (TC). Data split follows~\cite{shaker2022unetr++}. \textbf{Left Atrial (LA)}~\cite{xiong2021global} consists of 100 MRI scans with varying sizes, preprocessed using Z-score normalization~\cite{luo2021semi,wang2023mcf}. Five-fold cross-validation was used to segment the left atrium.  Evaluation metrics include DSC and HD95, with final segmentation generated using 0.5 patch overlaps. All datasets used a batch size of 2. The model, implemented in PyTorch 2.1.0~\cite{paszke2019pytorch} and for fair comparison integrated into the nnUNet framework~\cite{isensee2021nnu}, was trained on two NVIDIA A100 GPUs (80 GB VRAM). Training used a composite loss (DICE and cross-entropy, weighted 1:1) and the SGD optimizer with Nesterov momentum (0.99) and a weight decay of \(3 \times 10^{-5}\). The learning rate followed a polynomial decay strategy (power 0.9), set to an initial learning rate of 0.01 for isometric patches and 0.003 otherwise. Data augmentation followed~\cite{zhou2021nnformer,shaker2022unetr++,isensee2021nnu}. Training ran for 1000 epochs with 250 iterations per epoch (250,000 iterations). \Cref{tab:experimets-config} details the network settings for each dataset.

\begin{figure*}[!t]
  \centering
  \setlength\tabcolsep{1pt} 
        \resizebox{.95\textwidth}{!}{
        \begin{tabular}{@{} c!{\vrule width 2pt}c @{}}
                \begin{tabular}{ccc}
                 \multicolumn{3}{c}{\textbf{Synapse}}\\
                \multicolumn{3}{c}{\includegraphics[width=0.6\textwidth]{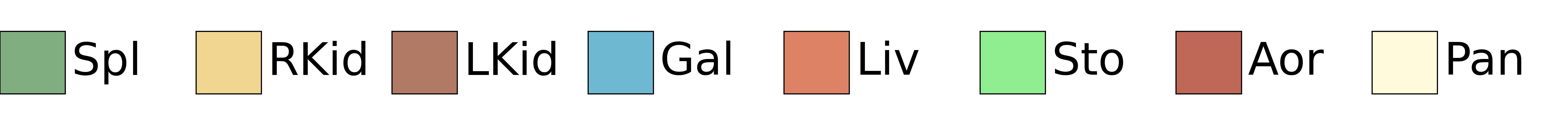}}\\
                  \includegraphics[width=0.2\textwidth]{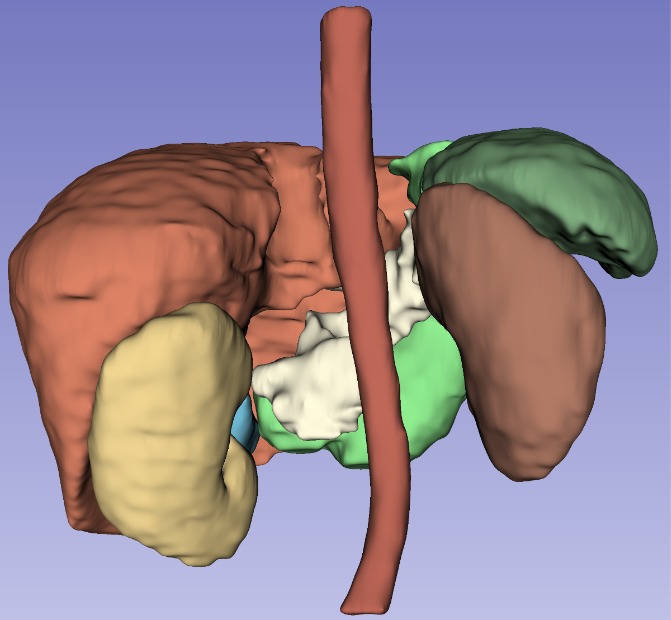} &
            \includegraphics[width=0.2\textwidth]{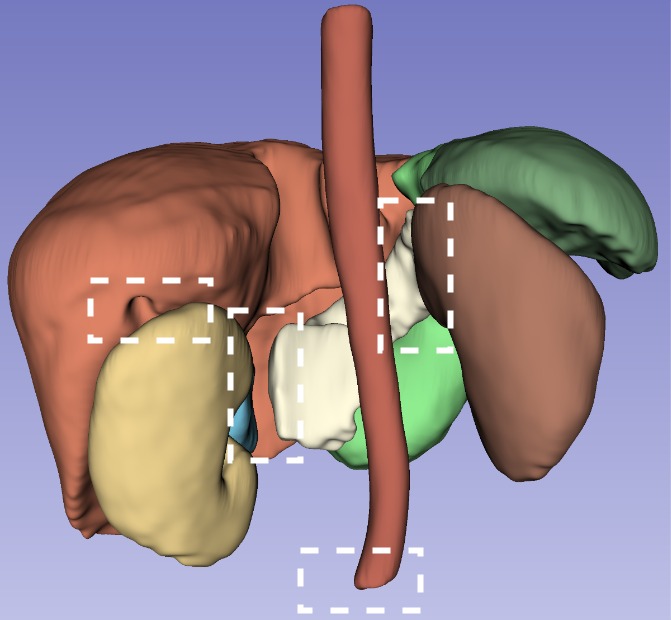} &
            \includegraphics[width=0.2\textwidth]{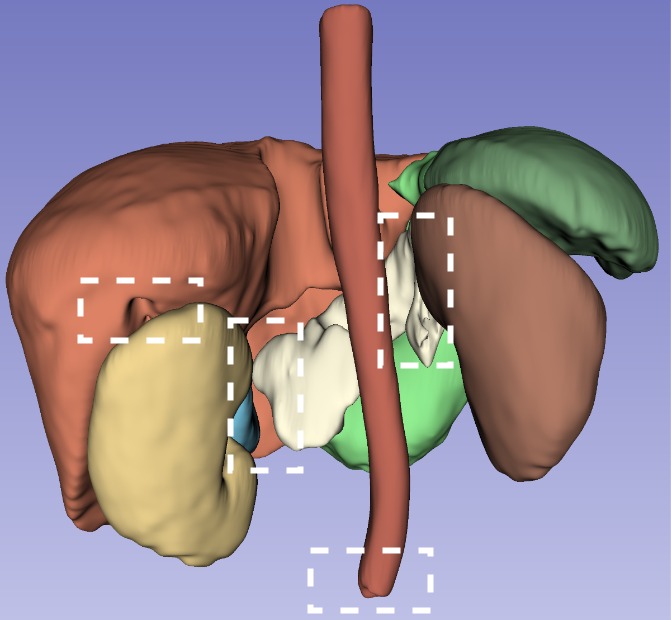} 
                \end{tabular} &

                \begin{tabular}{ccc}
                \multicolumn{3}{c}{\textbf{BraTS-Decathlon}}\\
                  \multicolumn{3}{c}{\includegraphics[width=0.2\textwidth]{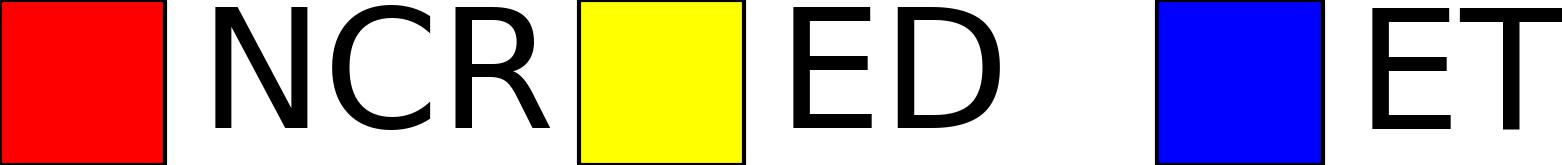}} \\
            \includegraphics[width=0.2\textwidth, height=0.135\textheight]{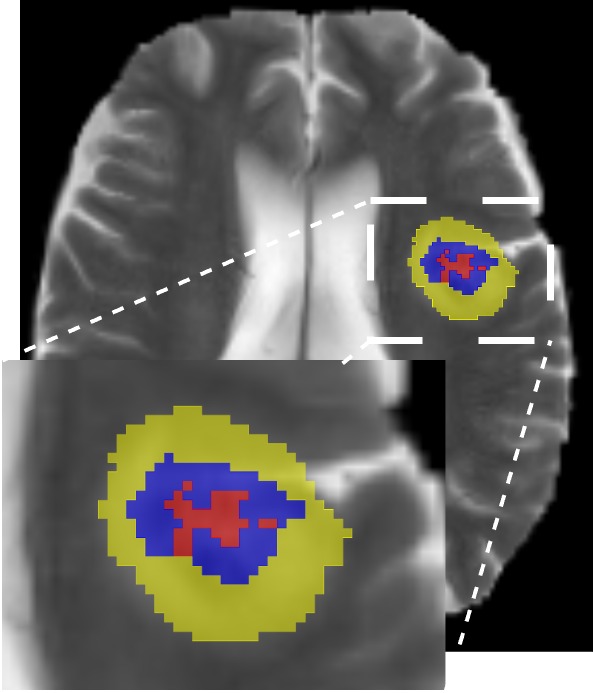} &
            \includegraphics[width=0.2\textwidth, height=0.135\textheight]{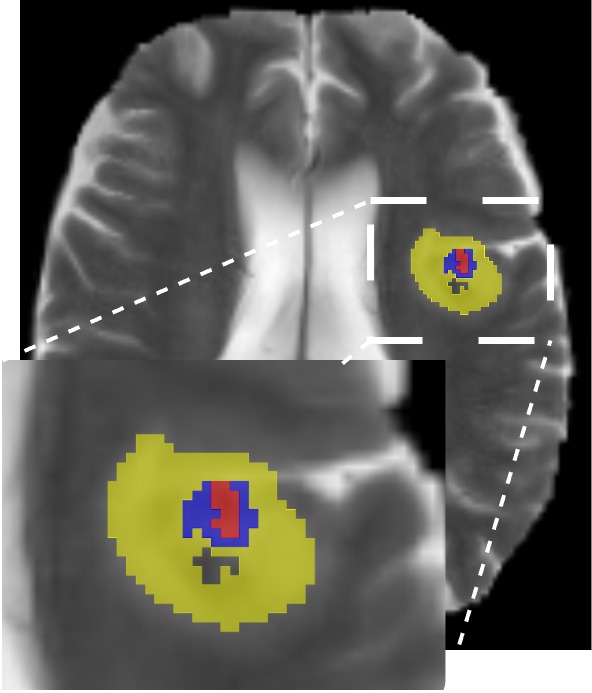} &
            \includegraphics[width=0.2\textwidth, height=0.135\textheight]{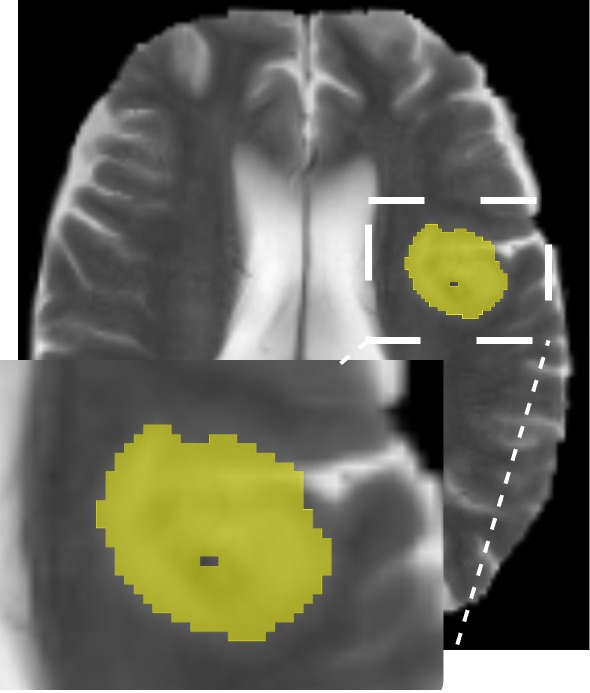} 
                \end{tabular} \\
                

                \begin{tabular}{ccc}
                \multicolumn{3}{c}{\textbf{Lung-Decathlon}}\\
                  \includegraphics[width=0.2\textwidth]{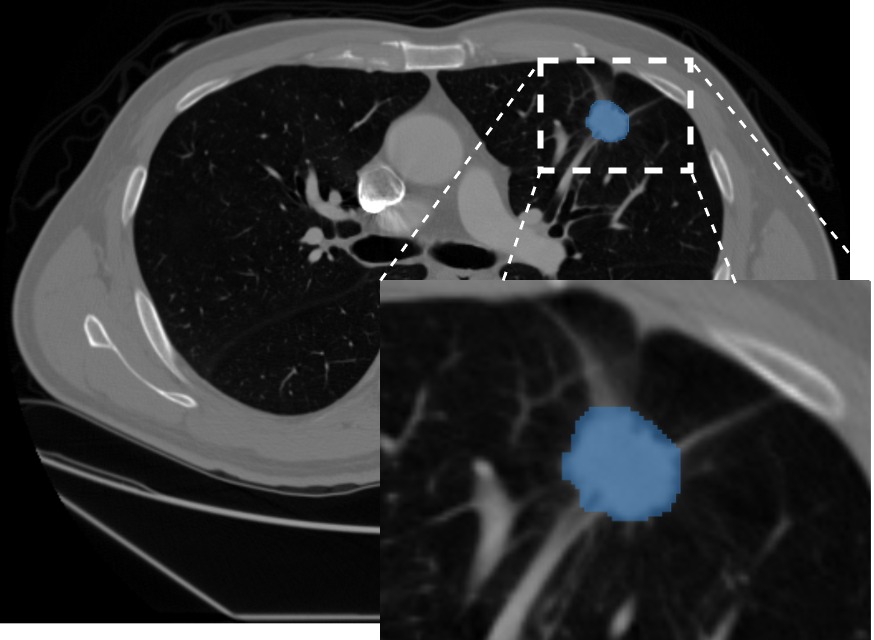} &
            \includegraphics[width=0.2\textwidth]{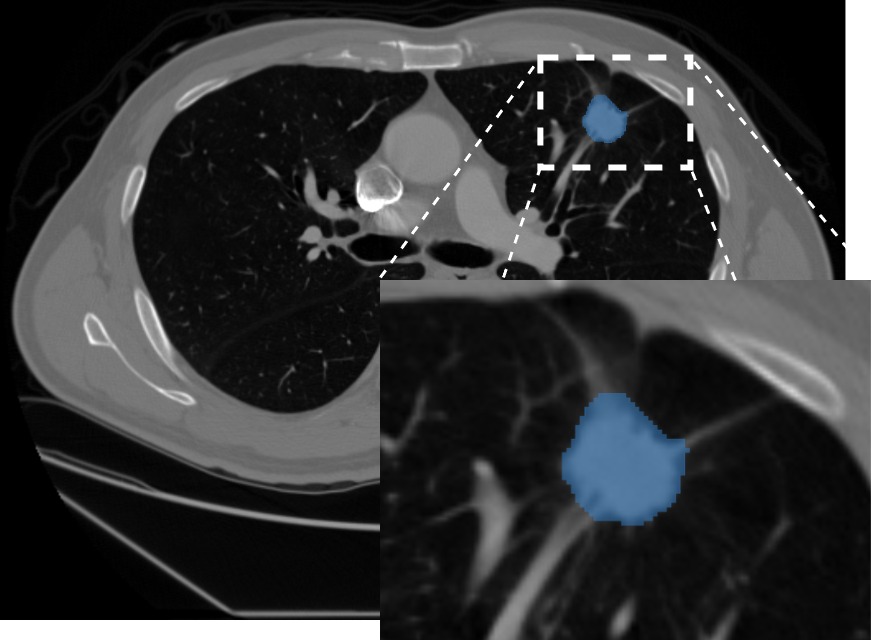} &
            \includegraphics[width=0.2\textwidth]{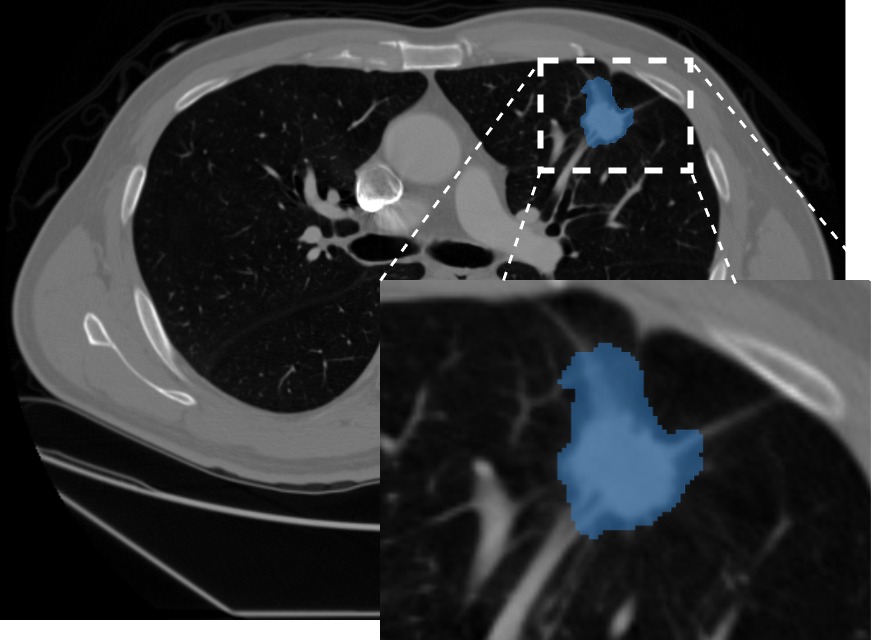}\\
                Groundtruth & LHU-Net & UNETR++
                \end{tabular} &

                \begin{tabular}{ccc}
                \multicolumn{3}{c}{\textbf{LA}}\\               
                  \includegraphics[width=0.2\textwidth, height=0.09\textheight]{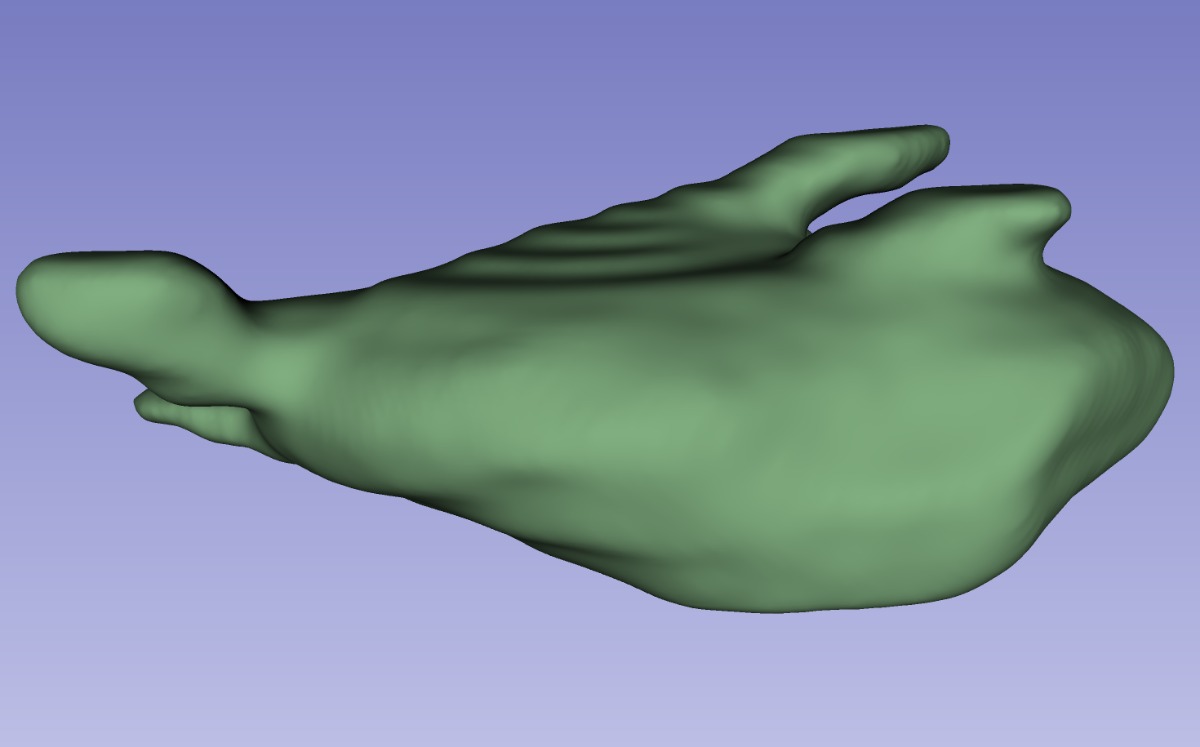} & \includegraphics[width=0.2\textwidth, height=0.09\textheight]{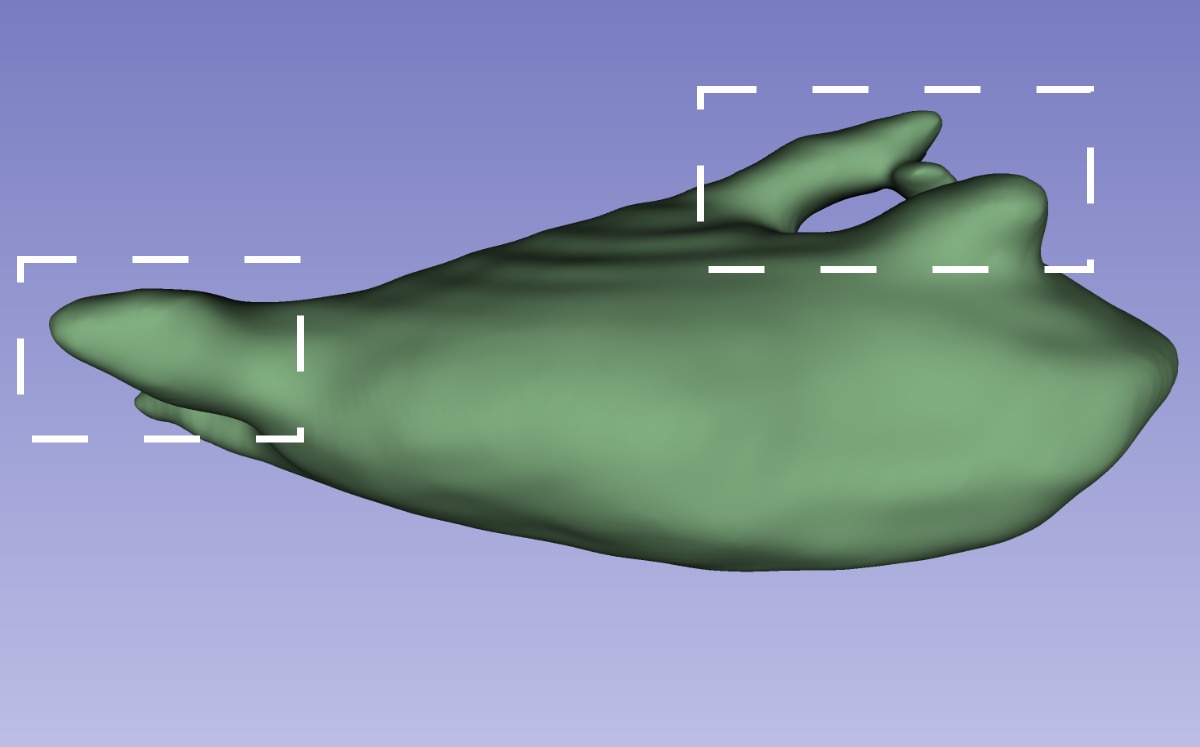} & \includegraphics[width=0.2\textwidth, height=0.09\textheight]{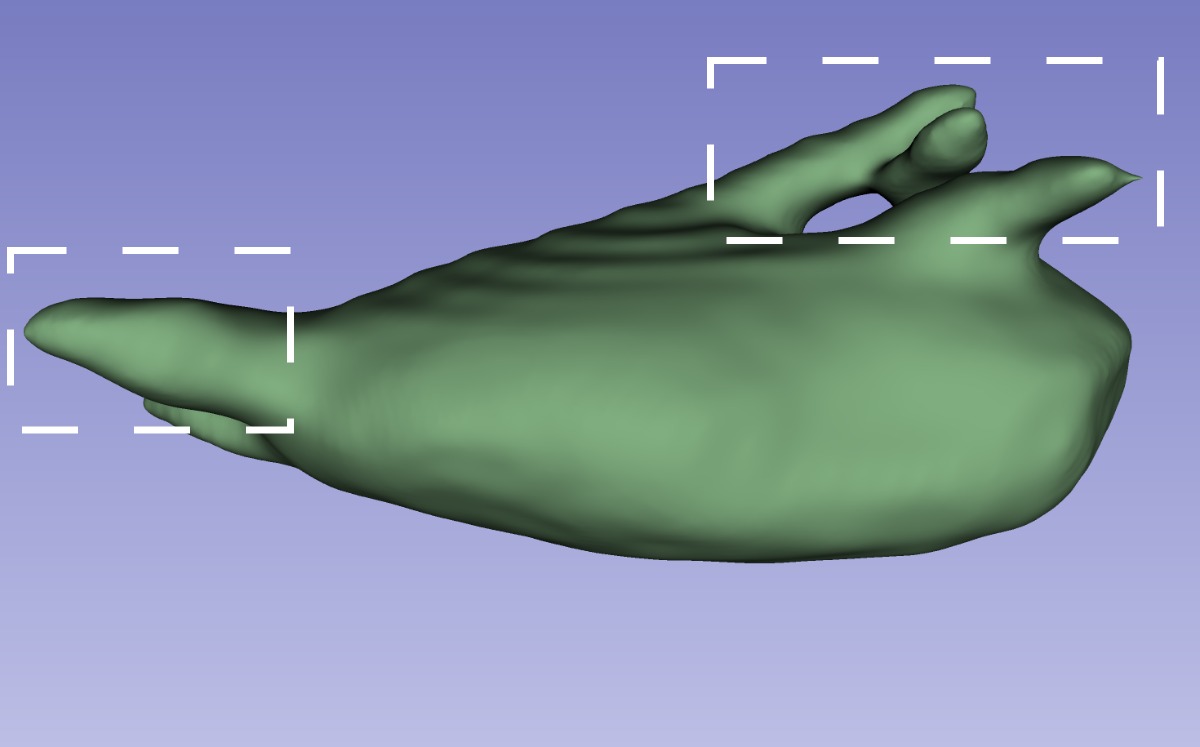} \\
                  Groundtruth & LHU-Net & UNETR++
                \end{tabular} \\
            
        \end{tabular}
    }
  \caption{Qualitative comparison of LHU-Net and UNETR++ across Synapse, BraTS-Decathlon, Lung-Decathlon, and LA datasets.}
  \label{fig:results-LA}
\end{figure*}
\subsection{Quantitative and qualitative results} \label{subsec:quantitative}

\Cref{tab:results-synapse} shows the performance of LHU-Net in comparison with SOTA models on the Synapse dataset. LHU-Net achieves the highest DSC while maintaining the lowest parameter count and FLOPs, offering a 4× reduction in parameters and 44\% lower FLOPs compared to the second-best DSC model. Additionally, LHU-Net improves average HD95 by 16\%, enhancing segmentation accuracy. Although MedNext~\cite{roy2023mednext} has a competitive parameter count, its DSC is 2\% lower than LHU-Net, with significantly higher FLOPs. \Cref{tab:results-LA} presents the quantitative results for the three datasets. As observed, LHU-Net surpasses other methods by a large margin. However, the HD95 in the Lung dataset reveals some shortcomings, indicating room for improvement. The parameters and FLOPs of each SOTA model (\Cref{tab:experimets-config}) highlight the efficiency of LHU-Net while outperforming SOTA models in Average DSC. \Cref{fig:results-LA} illustrates qualitative comparisons with UNETR++, one of the leading SOTA models. Across different datasets, UNETR++ exhibits over-segmentation or label omission, whereas LHU-Net consistently achieves more precise segmentation. This highlights how selecting the right module for each layer enhances efficiency while setting new benchmarks across datasets.

\section{Ablation studies}\label{sec:ablation}
We conducted an ablation study on the BraTS dataset to assess the impact of different attention mechanisms on segmentation performance. \Cref{tab:ablation-depth} presents the results, where attention mechanisms were applied to successive hybrid layers, including large kernel attention (L), LKAd (D), Self-Attention (S), Channel-Attention (C), and a baseline without CNN attention (I). The SSC-DDD configuration, which applies Self-attention (S) and channel attention (C) on the ViT side and LKAd (D) on the CNN side, achieves the highest DSC (83.81) with efficient computation. This demonstrates the advantage of selective attention combinations over uniform mechanisms. Reducing hybrid layers from three to two (SC-DD) slightly lowers DSC (82.85) but significantly reduces parameters (5.12M), indicating a strong trade-off between accuracy and efficiency. The III configuration (no CNN attention) performs worst (81.48 DSC), confirming the necessity of CNN attention mechanisms. Additionally, SCC-DDD achieves 83.12 DSC, suggesting that transitioning from self-attention (S) to channel attention (C) in earlier layers further enhances segmentation. Overall, large kernel convolutional attention followed by deformable convolution is crucial as a CNN attention in the hybrid layers, while progressive ViT attention transitions improve segmentation. Fewer hybrid layers can still offer competitive performance, making SC-DD a strong alternative for efficiency-focused applications.

The learnable weights (\(\gamma\) and \(\delta\)) that balance the contributions of CNN and ViT attention are crucial, as they vary across datasets (e.g., 0.45 for Synapse and 0.11 for BraTS). This demonstrates that fixed weights are suboptimal; learnable parameters allow the model to adapt and find the most suitable weight for each segmentation task.

\begin{table}[!t]
\renewcommand{\r}[1]{\textcolor{red}{#1}}
\renewcommand{\b}[1]{\textcolor{blue}{#1}}
    \centering
    \caption{Ablation study on the impact of different attention mechanisms on parameters, FLOPs, and DSC for the BraTS dataset. Each repeated entry represents attention used in successive hybrid layers. The best model is in \textbf{bold}.} 
    \resizebox{\textwidth}{!}{
    \begin{tabular}{c|c|c|c|c!{\vrule width 2pt}c|c|c|c|c}
    \toprule
    \textbf{ViT Attn.} & \textbf{CNN Attn.} & \textbf{Params.}\(\downarrow\) & \textbf{FLOPs}\(\downarrow\)  & \textbf{DSC}\(\uparrow\) & \textbf{ViT Attn.} & \textbf{CNN Attn.} & \textbf{Params.}\(\downarrow\) & \textbf{FLOPs}\(\downarrow\)  & \textbf{DSC}\(\uparrow\) \\
    \midrule
    SSS & DDD & 10.51 M & 57.44 G & 82.64 & SSC & DDI & 9.44 M & 57.33 G & 82.35 \\
    CCC & DDD & 7.97 M & 57.25 G & 82.52 & SSC & III & 8.88 M & 55.12 G & 81.48 \\
    \textbf{SSC} & \textbf{DDD} & \textbf{10.48 M} &\textbf{57.43 G}&\textbf{83.81} & SSC & LLL & 9.02 M & 55.55 G & 82.32 \\
    SCC & DDD & 10.33 M & 57.41 G & 83.12 & SC & DD & 5.12 M & 56.78 G & 82.85 \\
    \bottomrule
    \end{tabular}
    }
\label{tab:ablation-depth}
\end{table}


\section{Conclusion}\label{sec:conclusion}
In this work, we introduced LHU-Net, a lean hybrid U-Net for volumetric medical image segmentation. By strategically using spatial attention in early layers and channel attention in deeper layers, LHU-Net efficiently handles diverse datasets and segmentation tasks with only about 11 million parameters. Our results demonstrate that high segmentation accuracy can be achieved with a simpler model, advancing the development of accessible and effective medical image analysis tools.

\bibliographystyle{splncs04}
\bibliography{egbib}

\end{document}